\documentclass[]{spie}  

\usepackage{amsmath,amsfonts,amssymb}
\usepackage{graphicx}
\usepackage[colorlinks=true, allcolors=blue]{hyperref}
\usepackage[normalem]{ulem}

\title{Geant4 simulation of the residual background in the ATHENA Wide Field Imager from protons deflected by the Charged Particle Diverter}

\author[a,b]{G\'abor Galg\'oczi}
\author[a]{Jean-Paul Breuer}
\author[c]{Valentina Fioretti}
\author[d]{Jakub Zl\'amal}
\author[a]{Norbert Werner}
\author[d]{Vojtěch Čalkovský}
\author[e]{Nathalie Boudin}
\author[e]{Ivo Ferreira}
\author[e]{Matteo Guainazzi}
\author[f]{Andreas von Kienlin}
\author[g]{Simone Lotti}
\author[i]{Teresa Mineo}
\author[g]{Silvano Molendi}
\author[j]{Emanuele Perinati}
\affil[a]{Department of Theoretical Physics and Astrophysics, Faculty of Science, Masaryk University, Brno, Czech Republic}
\affil[b]{Institute of Physics, E\"otv\"os Lor\'and University, Budapest, Hungary}
\affil[c]{INAF/Osservatorio di Astrofisica e Scienza dello Spazio, Bologna, Italy}
\affil[d]{CEITEC, Brno University of Technology, Brno, Czech Republic}
\affil[e]{ESTEC/ESA, Keplerlaan 1, 2201AZ Noordwijk, The Netherlands}
\affil[f]{Max Planck Institut f\"ur extraterrestrische Physik, Garching bei M\"unchen, Germany}
\affil[g]{INAF/Istituto di Astrofisica Spaziale e Fisica cosmica, Milan, Italy}
\affil[i]{INAF/Istituto di Astrofisica Spaziale e Fisica Cosmica, Palermo, Italy}
\affil[j]{Eberhard Karls University, T\"ubingen, Germany}

\authorinfo{Further author information: (Send correspondence Galg\'oczi)\\  E-mail: gaborgalgoczi@gmail.com}

\begin{document} 
\maketitle

\begin{abstract}

X-ray telescopes opened up a new window into the high-energy universe. However, the last generation of these telescopes encountered an unexpected problem: their optics focused not only X-rays but low-energy (so called soft) protons as well. These protons are very hard to model and can not be distinguished from X-rays. For example, 40\% of XMM-Newton observations is significantly contaminated by soft proton induced background flares. In order to minimize the background from such low-energy protons the Advanced Telescope for High ENergy Astrophysics (ATHENA) satellite introduced a novel concept, the so called Charged Particle Diverter (CPD). It is an array of magnets in a Hallbach design, which deflects protons below 76 keV before they would hit the Wide Field Imager (WFI) detector. In this work, we investigate the effect of scattering of the deflected protons with the CPD walls and the inner surfaces of the WFI detector assembly. Such scattered protons can loose energy, change direction and still hit the WFI. In order to adopt the most realistic instrument model, we imported the CAD model of both the CPD and the WFI focal plane assembly. Soft protons corresponding to $\approx$2.5 hours of exposure to the L1 solar wind are simulated in this work. The inhomogeneous magnetic field of the CPD is included in the simulation. We present a preliminary estimate of the WFI residual background induced by soft proton secondary scattering, in the case of the optical blocking filter present in the field of view. A first investigation of the volumes responsible for scattering the protons back into the field of view is reported.


\end{abstract}


\pagebreak

\section{Introduction}
\label{sec:intro}  

Observations in the X-ray spectral band are vital for our understanding of the high-energy Universe, such as supernova remnants, galaxies and clusters of galaxies. The Einstein Observatory (also named HEAO-2, flown in 1978–1981) was the first X-ray telescope that utilized the Wolter Type I grazing-incidence X-ray optics that most observatories use to focus X-rays. Two of the most famous and renowned telescopes are NASA's Chandra X-ray Observatory (1999-) and the European Space Agency's (ESA) X-ray Multi-Mirror Mission Newton (1999-). The previous one focused on having superior spatial resolution on a limited field of view by optimizing the angular resolution at the cost of mass and effective area. The optics of the XMM-Newton on the other hand is based on electro-formed Nickle, replicated from precision mandrels. This maximised the effective area to be able to offer spectral information on far fainter objects. It was an unexpected development that low-energy protons ($\le$300 keV) can scatter on the optics and end up being detected. 



\begin{figure} [!h]
   \begin{center}
   \begin{tabular}{c} 
   \includegraphics[width=0.7\textwidth]{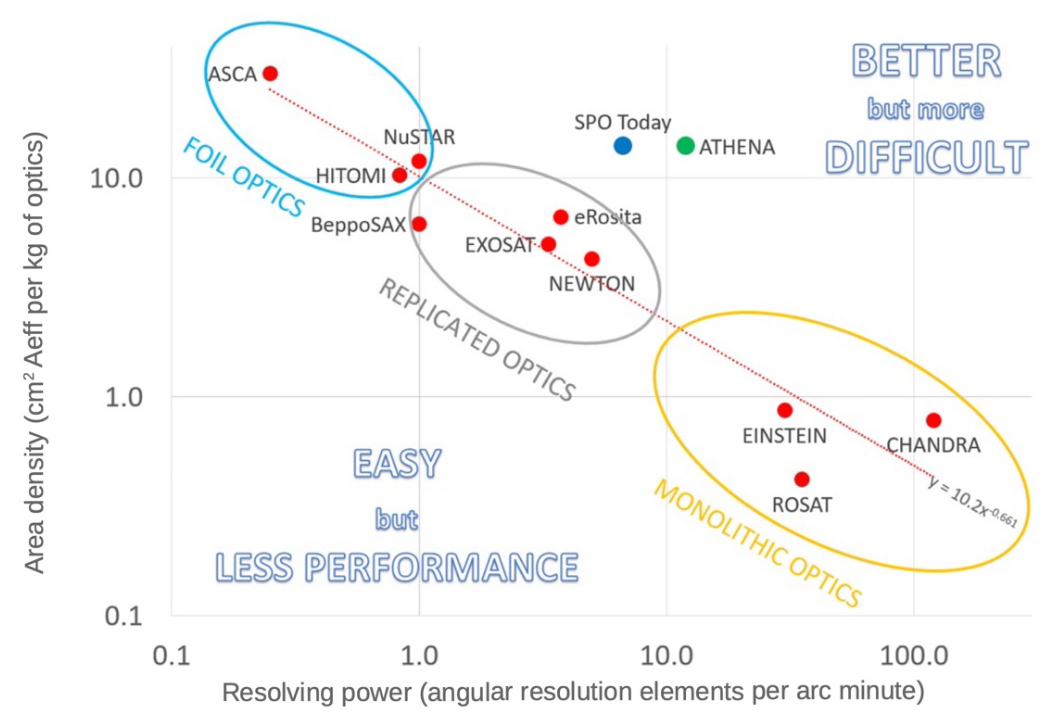}
	\end{tabular}
	\end{center}
   \caption{ Classifying past and existing X-ray telescopes and the ATHENA telescope by area density of the optics and resolving power. ATHENA is an outlier, since it excels in both parameters.
   \cite{firstfig}
   \label{fig:athena}}
\end{figure}

The next generation high-energy astrophysics observatories will look deeper into the high-energy Universe. Two metrics that are often used to classify X-ray telescopes are the area density and resolving power. The next large-class telescope of ESA will be the Advanced Telescope for High-ENergy Astrophysics (ATHENA). In Fig.~\ref{fig:athena} several of the past X-ray telescopes are compared to ATHENA with the mentioned two metrics. ATHENA will have an effective area of 1.4 m$^2$ at 1 keV with a focal length of 12 m. In order to obtain superior area density with fair resolving power ATHENA will utilize the so called Silicon Pore Optics (SPO) using a Wolter-I system. Interestingly, its design and production is based on the processes of the semiconductor industry. Since there have been tremendous efforts in producing electronics in the industry, this know how enables the production of an optical system that has excellent X-ray focusing capabilities \cite{myspo}. 

The planned position of the ATHENA satellite is at the first Lagrange point as it is more favourable for the background \cite{valaas,L1}. The two instruments of ATHENA are the Wide Field Imager (WFI) and X-ray Integral Field Unit (X-IFU). This work focuses on the soft proton induced background of the WFI, sensitive in the 0.2~-~15 keV energy band. The origin of such low-energy ($\le$300 keV) protons is an actively studied field. Their possible sources include interplanetary
coronal mass ejections, acceleration events in Earth’s magnetosphere and flaring near open coronal field lines \cite{softprotonorigin}. These protons can scatter through the optical system of ATHENA and generate counts similar to the investigated X-rays in the detectors, resulting in contaminated measurements. Previous simulations predict the WFI soft proton induced background without magnetic shielding would be order of magnitudes higher than the requirement \cite{valaas}.


The ESA and ATHENA team developed a novel concept to minimize the background induced by soft protons\cite{cpd}. A Hallbach array of magnets is placed between the optical system and each of the detectors. The one in front of WFI is pear-shaped, has a diameter of 35 cm and its magnets weigh 21 kg. The requirement for the maximum residual background for the 2-7 keV band from soft protons is $5\times 10^{-4}$ cts/cm$^2$/keV/s, 10 \% of the non-focused non X-ray background requirement\cite{wfibkg}.

The scientific assessment of this Charged Particle Diverter (CPD) is ongoing under the Background Topical Panel of the ATHENA Science Study Team \cite{sciass}, that includes the end-to-end simulation of the proton transmission efficiency from the SPO to the focal plane, using independent SRIM \cite{jakub} and Geant4 simulations. Within this activity, we simulated in Geant4 the deflection of the WFI CPD and the potential secondary scattering of the deflected protons with the CPD walls, the WFI baffle and the other volumes composing the assembly. The final aim is estimating the residual background on the WFI, comparing it to the requirement and minimize, if needed, the secondary scattering in the WFI system.

Section 2.1 details the geometry applied in Geant4. Section 2.2 describes how the magnetic field of the CPD was treated in Geant4 and how the effect of the optics in front of WFI was taken into account. Section 3. describes the deposited energy spectrum on the WFI. 

\section{Simulating the proton scattering in GEANT4}

A dedicated simulation was developed to understand if protons after being deflected by the CPD can scatter on the inner surfaces of the focal plane assembly and contribute to the background of the ATHENA WFI detector. The simulation was based on the Geant4 toolkit library \cite{geant2,geant1}. All simulations in this work were based on Geant4 version 10.5 with the patch 01 and use the single Coulomb scattering model as defined in the "G4EmStandardPhysicsSS" electromagnetic physics list \cite{singlescatmodel}. The CAD models utilized in the simulation were imported to Geant4 directly with CADMesh \cite{cadmesh}. 

\subsection{The geometry of the simulation}

\begin{figure} [!h]
   \begin{center}
   \begin{tabular}{c} 
   \includegraphics[width=0.8\textwidth]{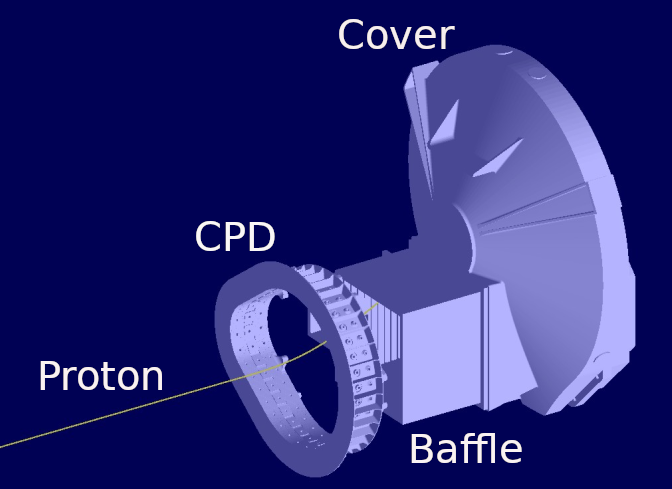}
	\end{tabular}
	\end{center}
   \caption{The model of the ATHENA WFI detector system and the Charged Particle Diverter imported in Geant4, kindly provided by the WFI team and ESA. A proton that is deflected by the magnetic field of the CPD is visible hitting the baffle volume. Three of the six included CAD volumes are visible. \label{fig:cad}}
\end{figure}

\begin{figure} [!h]
   \begin{center}
   \begin{tabular}{c} 
   \includegraphics[width=0.8\textwidth]{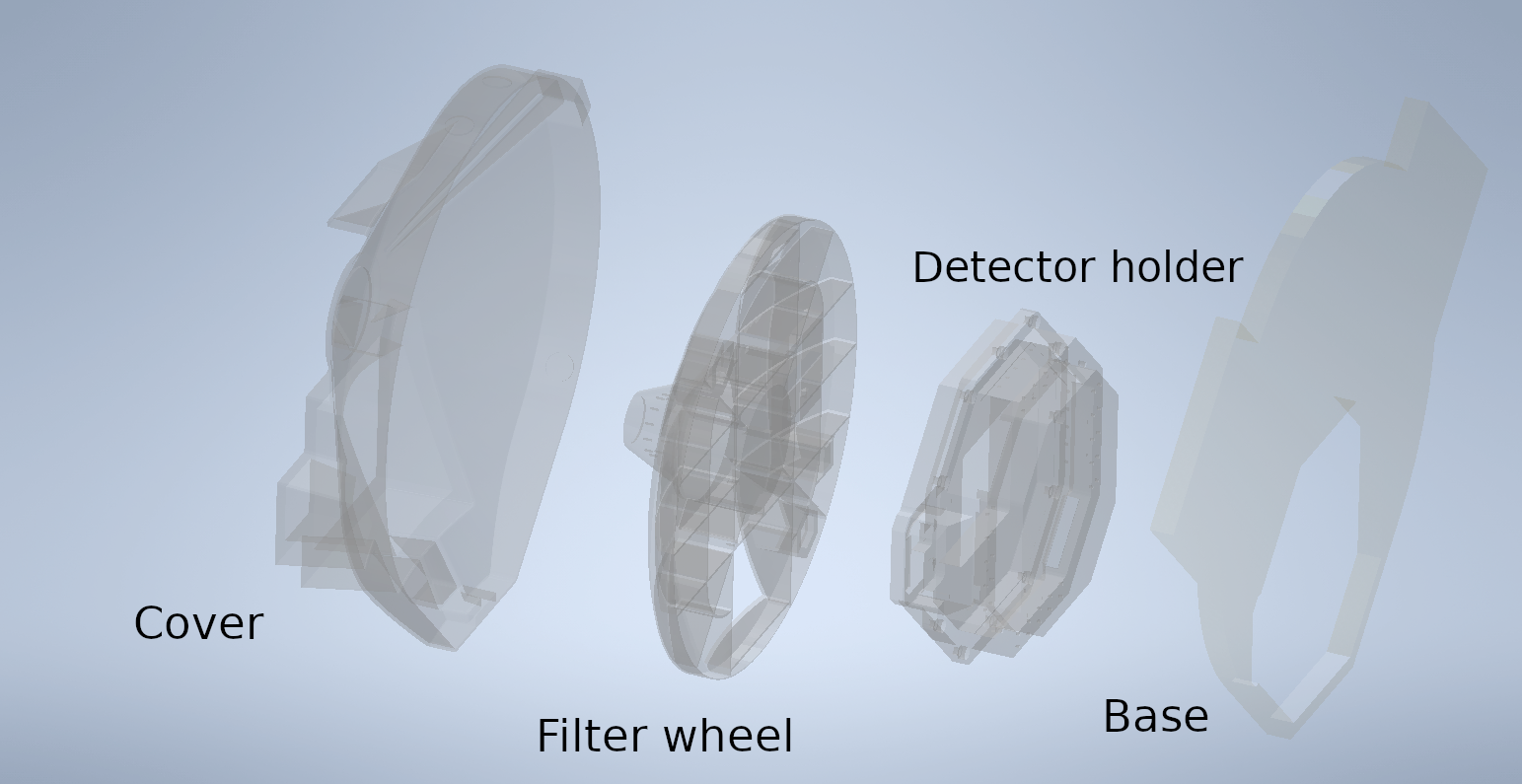}
	\end{tabular}
	\end{center}
   \caption{Exploded view of the CAD volumes under the "cover" volume. Protons arrive from the left side, through the hole in the cover. Then they traverse the filter wheel and its optical blocking filter. Finally, they arrive at the detector, which is situated inside the detector holder.   \label{fig:cad2}}
\end{figure}

The inner part of the ATHENA telescope, which contains the WFI detector, has eight separate volumes. The incoming protons cross the CPD and its magnetic field first as can be seen in Fig.~\ref{fig:cad}. Some of the protons hit the walls of the CPD.  Protons then pass through (or scatter) on the baffle. Inside the cover volume there are five volumes. Three of them being CAD volumes, which can be seen in
Fig.~\ref{fig:cad2}: the filter wheel, the holder of the detector and the base of the system. The remaining two volumes are the detector itself and the optical blocking filter which sits inside the filter wheel. By importing CAD models \cite{jakub}, we both saved time and made it possible to include small structures, such as the ridges on the surface of the baffle and the saw shape walls of the pear-shaped CPD. It is important to mention that these volumes are assumed flat on the micrometer scale, which is not the case in reality.

The WFI baffle has a coating of "Magic Black" material that consists of 44\% Oxygen and 56\% Aluminium by weight. The composition was measured by EDS X-Max 20 by Oxford Instruments installed in Scanning Electron Microscope TESCAN Mira on BUT CEITEC Nano (https://nano.ceitec.cz/scanning-electron-microscope-e-beam-writer-tescan-mira3raith-lis-mira/). CPD is coated with "Surtec 650" material (details in the Finitec document reference number: 1558830-914\_00). The rest of the CAD volumes have the so called "AL7075" alloy, which contains Aluminium, 5.8\% Zinc, 2.3\%, Magnesium, and 1.3\% Copper by weight. The Geant4 mass models of the WFI large area detector and the optical blocking filter are based on the work of Fioretti et al. (2018)\cite{valaas}.

\pagebreak

\pagebreak

\subsection{Input protons and magnetic field of the simulation}


A dedicated raytracing simulation of the proton transmission through the ATHENA Silicon Pore Optics was performed under the CPD scientific assessment \cite{sciass}, using the Remizovich solution as scattering model \cite{softscatssmodel} and assuming the L1 proton solar wind model as input population. The protons reaching the CPD were used as input for the present simulation. The list included 62.5 million protons, which corresponded to 2.4 hours of time in orbit. All of these were simulated one by one.

\begin{figure} [!h]
   \begin{center}
   \begin{tabular}{c} 
   \includegraphics[width=1\textwidth]{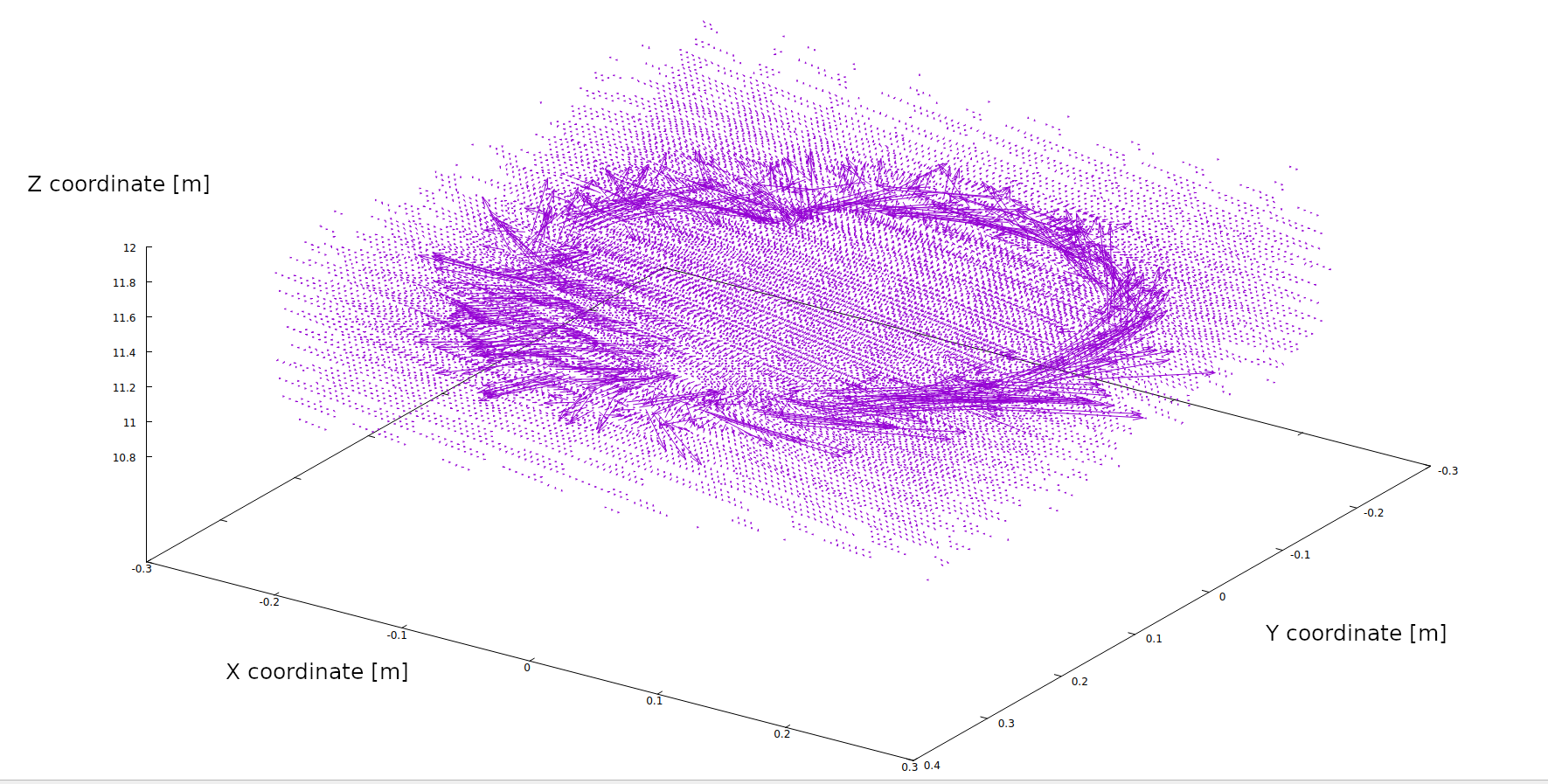}
	\end{tabular}
	\end{center}
   \caption{ The magnetic vector field of the CPD used in the Geant4 simulation. The CPD consists of many smaller magnets in a pear-shaped configuration, producing the inhomogeneous field. \label{fig:mag}}
\end{figure}

The magnetic field of the CPD was calculated with the Finite Element Method and was read into our particle simulation. In Geant4 one can customize the parameters of the trajectory equation solver. In our case, we chose the fourth order Runge–Kutta solver with six number of variables. The minimal step in the field was chosen to be 0.1 mm. The miss distance, which describes how curved the paths of charged particles are approximated -- by a set of linear chord segments -- was chosen to be 1 $\mu$m. The "delta intersection" parameter defines the minimum accuracy for the distance of track with a volume boundary. It was set to 0.1 mm. "Delta onestep" is coupled to "delta intersection", it was set to be 1 $\mu$m. The relative error for the integration of the steps was also set. Its minimum value was chosen to be 10$^{-5}$ and the maximum to 10$^{-4}$. Several other configurations were investigated but these settings proved to be the fastest and most accurate. Fig.~\ref{fig:mag} shows the magnetic field used in the present study. For every thousandth point in the grid the magnetic field vector is plotted. Since the CPD consists of a set of magnets in a pear-shape, the field is very inhomogeneous. The strongest field on the grid is 1.56 T (inside the magnet), while the weakest one is 136 $\mu$T (far from the CPD). Z direction is the longitudinal axis of the satellite.

The requirement for the magnetic field of the diverter was to deflect "direct" protons with energies below 76 keV before they reach the WFI detector. In our case, the magnetic field was sampled on a Cartesian grid with a step size of 7.5 mm and a volume of 0.125 m$^3$ surrounding the CPD in the simulation. The list of the magnetic field vectors included about 2.5 million points. This was read into the randomly accessible memory. A dedicated C++ package was developed to transform these points onto an other Cartesian grid with one unit cubes. This way the magnetic field for any given coordinate could be retrieved by its index speeding up the Geant4 simulation a thousand-fold. The paths of individual proton tracks were validated in our simulation with an independent COMSOL Multiphysics simulation.

\pagebreak

\section{Results}

The $\approx$one million pixels of the WFI detector registered how much energy was deposited in them. The energy of the primary proton corresponding to each hit was also registered. We have disregarded events where more than one pixel was hit to simplify the pattern reconstruction of the events, since multi-pixel events have less than 1\% occurrence. 



\begin{figure} [!h]
   \begin{center}
   \begin{tabular}{c} 
   \includegraphics[width=1\textwidth]{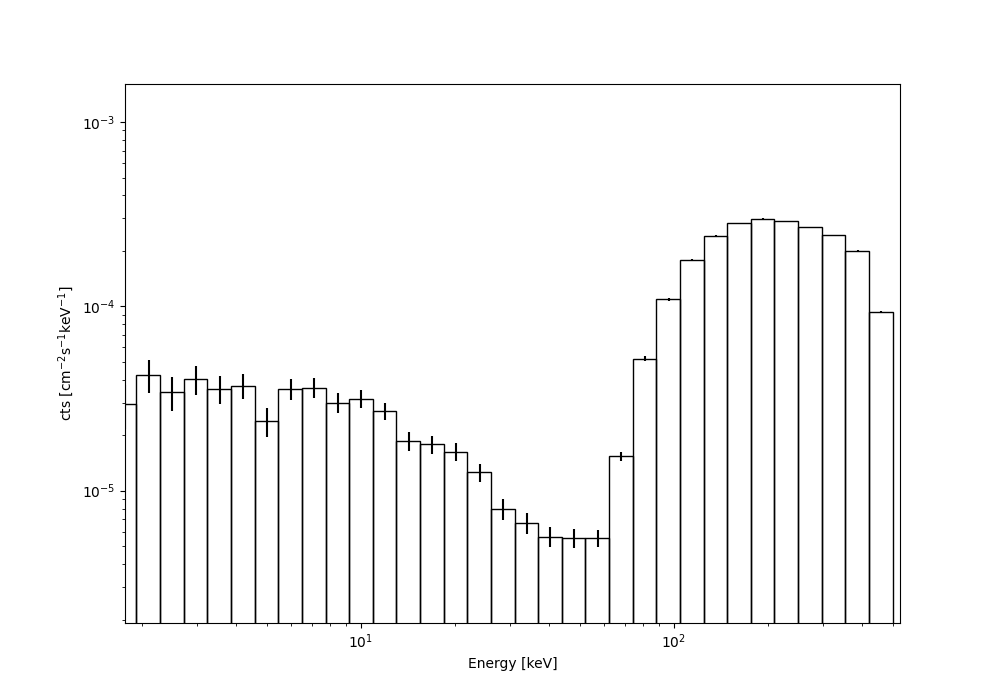}
	\end{tabular}
	\end{center}
   \caption{Residual background for the ATHENA WFI. Even though the initial proton energy follows a power law, the protons with lower energy are suppressed as they are deflected by the magnetic field of the CPD.
   \label{fig:edep}}
\end{figure}

In Fig.~\ref{fig:edep} the energy deposition by soft protons in the ATHENA WFI detector is plotted. 
The background in the most relevant energy band (2-7 keV) was found to be 3.5 ± 0.2 × 10$^{-5}$ cm$^{-2}$s$^{-1}$keV$^{-1}$. This value includes the scattered protons. It is more than an order of magnitude lower than the actual requirement for the WFI detector (5 × 10$^{-4}$ cm$^{-2}$s$^{-1}$keV$^{-1}$) [2]. 

In order to understand which volumes contribute to the background the most, we have registered each of them as the primary proton passed through. All together there were 279 protons that induced an event in the 2-7~keV energy band. 41 \% of them were scattered on the volume which holds the detector and 44 \% of them on the holder of the filter wheel. The rest of the volumes had negligible contribution as the third one by value was the baffle with 8 \%. 
Since we used a simplified model for the inner structures surrounding the WFI, further studies including a more detailed model of the detector holder are planned to verify their contribution to secondary scattering.


\section{Conclusion and final remarks}

A dedicated Geant4 simulation of soft protons entering the WFI detector system of the ATHENA telescope was performed. We proved that the magnetic field designed to deflect protons below $76$ keV has the expected effect. We included all the volumes composing the telescope assembly around the WFI detector and we showed that even though the expected residual background increased due to protons scattering on the inner surfaces, its level is still more than an order of magnitude lower than the requirement if the optical blocking filter is present. We also presented the first results on the investigation of volumes responsible for scattering back soft protons into the field of view of the WFI detector.

The simulation of the residual background without the use of the optical blocking filter is planned, together with the use of a more detailed model of the WFI inner structure. 

On the micrometer scale flat surfaces were assumed for all of the CAD models. It is important to mention that an undergoing investigation shows that surface roughness
can increase the scattering efficiency for such low energy protons. Although preliminary studies show that this increase has a negligible impact on the residual background, a dedicated study is ongoing~\cite{jp} to include the surface roughness in the Geant4 CPD simulation. 

\acknowledgments
We thank ESA and Brno University of Technology for kindly providing the CPD CAD model and the magnetic field density map. We thank the WFI team for kindly providing the WFI assembly CAD model.

JPB and NW are supported by the GACR grant 21-13491X. Part of the funding was provided by the Internal Grant Agency of Masaryk University, specifically the Operational Programme Research, Development and Education within the framework of the implemented project IGA MU reg. No. CZ.02.2.69/0.0/0.0/19 073/0016943 as well as by the MASH3 grant. 






\pagebreak

\bibliography{report} 
\bibliographystyle{spiebib} 

\end{document}